\documentclass[aps,prl,10pt,twocolumn,superscriptaddress,balancelastpage,showpacs,reprint]{revtex4-1}
\usepackage{centernot}
\usepackage{graphicx}
\usepackage{amsmath}
\usepackage{times}
\usepackage{amssymb}
\usepackage{mathrsfs}
\usepackage{chemarr}
\usepackage{color}
\usepackage{url}
\usepackage{version}
\usepackage[pdftex,colorlinks=true,
pdfstartview=FitV,
linkcolor= linkcolor,
citecolor= linkcolor,
urlcolor= linkcolor,
hyperindex=true,
hyperfigures=false]
{hyperref}

\definecolor{linkcolor}{rgb}{0,0,0.6} 

\newcommand{\bp}{{\bf p}}

\newcommand{\br}{{\bf r}}
\newcommand{\bv}{{\bf v}}

\newcommand{\cO}{\mathcal{O}}
\newcommand{\cP}{\mathcal{P}}

\newcommand{\dd}{\text{d}}
\newcommand{\ee}{\text{e}}

\newcommand{\B}{\text{\tiny B}}

\newcommand{\R}{\text{\tiny R}}

\providecommand{\avg}[1]{\left \langle #1 \right \rangle}
\providecommand{\pnt}[1]{\left  ( #1 \right ) }
\providecommand{\brt}[1]{\left  [ #1 \right ] }

\providecommand{\abs}[1]{\left | #1 \right|}
\providecommand{\f}[2]{\frac{ #1}{#2}}

\begin{document}


\title{How far from equilibrium is active matter?}

\author{\'Etienne Fodor}
\affiliation{Universit\'e Paris Diderot, Sorbonne Paris Cit\'e, MSC, UMR 7057 CNRS, 75205 Paris, France}
\author{Cesare Nardini}
\affiliation{SUPA, School of Physics and Astronomy, University of Edinburgh, Edinburgh EH9 3FD, United Kingdom}
\affiliation{DAMTP, Centre for Mathematical Sciences, University of Cambridge,
Wilberforce Road, Cambridge CB3 0WA, United Kingdom}
\author{Mike E. Cates}
\affiliation{SUPA, School of Physics and Astronomy, University of Edinburgh, Edinburgh EH9 3FD, United Kingdom}
\affiliation{DAMTP, Centre for Mathematical Sciences, University of Cambridge,
Wilberforce Road, Cambridge CB3 0WA, United Kingdom}
\author{Julien Tailleur}
\affiliation{Universit\'e Paris Diderot, Sorbonne Paris Cit\'e, MSC, UMR 7057 CNRS, 75205 Paris, France}
\author{Paolo Visco}
\affiliation{Universit\'e Paris Diderot, Sorbonne Paris Cit\'e, MSC, UMR 7057 CNRS, 75205 Paris, France}
\author{Fr\'ed\'eric van Wijland}
\affiliation{Universit\'e Paris Diderot, Sorbonne Paris Cit\'e, MSC, UMR 7057 CNRS, 75205 Paris, France}

\date{\today} \pacs{}

\begin{abstract}
  Active matter systems are driven out of thermal equilibrium by a
  lack of generalized Stokes-Einstein relation between injection and
  dissipation of energy at the microscopic scale. We consider such a
  system of interacting particles, propelled by persistent noises, and
  show that, at small but finite persistence time, their dynamics
  still satisfy a time-reversal symmetry. To do so, we compute
  perturbatively their steady-state measure and show that, for short
  persistent times, the entropy production rate vanishes. This endows
  such systems with an effective Fluctuation-Dissipation theorem akin
  to that of thermal equilibrium systems. Last we show how interacting
  particle systems with viscous drags and correlated noises can be
  seen as in equilibrium with a visco-elastic bath but driven out of
  equilibrium by non-conservative forces, hence providing an energetic
  insight on the departure of active systems from equilibrium.
\end{abstract}

\maketitle 

Active matter systems comprise large assemblies of individual units
that dissipate energy, often stored in the environment, to produce
mechanical work~\cite{Marchetti2013RMP}. From the collective motion of
self-propelled particles~\cite{Vicsek1995PRL,Vicsek2012PR} to the
existence of a liquid phase in the absence of attractive
forces~\cite{Tailleur:08,Fily:12,Cates:15} many intriguing phenomena
have generated a continuously growing interest for active matter over
the past decades~\cite{Marchetti2013RMP}. Since active systems break
detailed balance at the microscopic scale, they cannot be described by
equilibrium statistical mechanics. However, it is often difficult to
pinpoint precisely the signature of non-equilibrium physics in their
emerging properties. For instance, motility-induced phase separation,
which leads to the liquid-gas coexistence of repulsive self-propelled
particles, is not associated to the emergence of steady-state mass
currents. A number of works have actually proposed that its large
scale physics can be captured by an equilibrium
theory~\cite{Tailleur:08,Speck2014PRL,Takatori2014PRL,Brader:15}, the
limits of which are heavily
debated~\cite{wittkowski2014scalar,Solon:15a,Bialke2015PRL}. Even for
systems where steady currents arise the connection to equilibrium
physics can sometimes be maintained, as for the transition to
collective motion which amounts, for simple systems, to a liquid-gas
phase transition~\cite{SolonPRL2013,solon2015phase}. More and more
approaches to active matter thus partly rely on the intuition built
for equilibrium
systems~\cite{Tailleur:08,Speck2014PRL,Takatori2014PRL,Mallory2014PRE,Yang2014PRL,Brady:15,Ginot2015PRX,Bialke2015PRL}.

Building a thermodynamic approach for active matter thus first require
understanding how active systems depart from thermal
equilibrium. Insight into this question was gained by studying how the
Fluctuation Dissipation Theorem (FDT) breaks down in active
matter~\cite{Loi2008PRE,Loi2011SM,Szamel:14,Solon:15c}. At short
time and space scales, the persistent motion of active particles
typically precludes the existence of effective temperatures while at
larger scales FDTs can sometime be recovered. In living systems, the
violation of FDT is used to characterize the forces generated by
intracellular active processes~\cite{Mizuno, Wilhelm, gallet:2009,
  benisaac:2011, Visco:15, Ahmed:15b}. The information extracted from
violations of the FDT is however rather limited and non-equilibrium
statistical mechanics offers more elaborate tools to quantify the
departure from equilibrium.  In particular, the entropy production
rate quantifies the breakdown of time-reversal symmetry, whence
probing the irreversibility of the dynamics~\cite{Lebowitz}. Hard to
compute, and even harder to measure experimentally, it has been little
studied in active systems~\cite{Chaudhuri:13,Chaudhuri:14}, hence the
need for `simple but not simpler' systems which offer a natural way to
establish theoretical frameworks.

In this letter we study a model system of active matter which has
recently attracted lots of
interest~\cite{Maggi:14b,Maggi:15b,Szamel:15,Brader:15}. It comprises
overdamped `self-propelled' particles whose dynamics read
\begin{equation}\label{eq:AOU} 
  \dot {\bf r}_i= -\mu \nabla_i \Phi + \bv_i ,
\end{equation}
where $i$ refers to the particle label, $\mu$ to their mobility and
$\Phi$ is an interaction potential. The self-propulsion velocities
${\bv_i}$, rather than having fixed norms as in models of Active
Brownian Particles~\cite{Fily:12} (ABPs) or Run-and-Tumble
Particles~\cite{Schnitzer:93} (RTPs), are zero-mean persistent
Gaussian noises of correlations $\avg{ v_{i\alpha}(t) v_{j\beta}(0) }
= \delta_{ij}\delta_{\alpha\beta} \Gamma(t)$, with greek indices
corresponding to spatial components. In the simplest of cases, the
$\bv_i$'s are Ornstein-Uhlenbeck processes, solutions of $\tau \dot
\bv_i=-\bv_i +\sqrt{2D} \boldsymbol{\eta}_i$, with
$\boldsymbol{\eta}_i$'s zero-mean unit-variance Gaussian white noises,
so that $\Gamma(t) = D \ee^{ - \abs{t} / \tau } / \tau$. Here $D$ controls
the amplitude of the noise and $\tau$ its persistence time.

Since the temporal correlations of the noise are not matched by
similar correlations for the drag, this system does not satisfy the
standard generalization of the Stokes-Einstein relation to systems
with memory~\cite{Kubo}. Consequently, the system is out of thermal
equilibrium and its stationary measure is not the Boltzmann weight
$P_\B \equiv Z^{-1} \exp(-\beta\Phi)$. This model, to which we refer
in the following as Active-Ornstein-Uhlenbeck Particles (AOUPs),
shares the essential features of active systems: it correctly
reproduces the behavior of passive tracers in bacterial
baths~\cite{Maggi:14a, Maggi:14b}, leads to the standard accumulation
of active particles close to confining walls~\cite{Maggi:15b}, and
shows a shifted onset of the glass transition~\cite{Szamel:15}.  As
for many other self-propelled particle
systems~\cite{Tailleur:09,Szamel:14}, the limit of vanishing
persistence time of AOUPs correspond to an equilibrium Brownian
dynamics, since ${\bf v}_i$ reduces to a Gaussian white noise.

In the following, we characterize how the AOUPs depart from thermal
equilibrium. First, we compute perturbatively their steady-state at
small but finite persistence time $\tau$. Surprisingly, we show that
the small $\tau$ limit yields a non-Boltzmann distribution with which
the system \textit{still respects detailed-balance}: The entropy
production, which we compute, can indeed be shown to vanish at order
$\tau$. In this regime, to which we refer as \textit{effective
  equilibrium}, we also show that AOUPs satisfy a generalized FDT.
Finally, we close this article by providing an energetic
interpretation of the breakdown of detailed-balance for AOUPs.

\begin{figure}
  \includegraphics[width=.45\columnwidth]{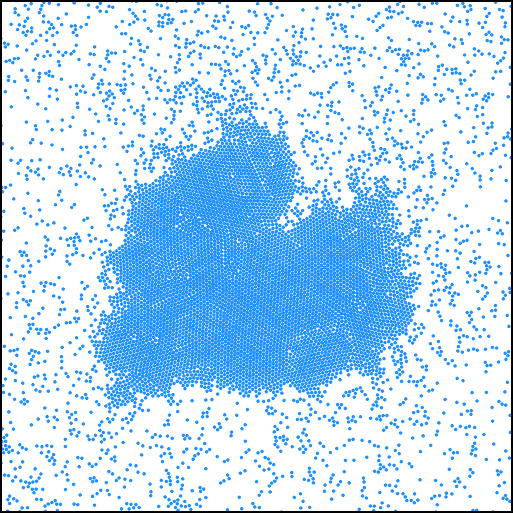}
  \caption{AOUPs interacting via the repulsive
    potential~\eqref{eq:repul} exhibit MIPS in a 2d box of size $L$
    with periodic boundary conditions. Parameters: $A=100,\,
    \sigma=2,\,N=10\,000,\, D=1000,\,\tau=20$}\label{fig:MIPS}
\end{figure}

We consider $N$ particles, propelled by Ornstein Uhlenbeck processes,
interacting through a potential $\Phi$. For illustration purposes, we
use pairwise repulsive forces in 2D
\begin{equation}\label{eq:repul}
  \Phi=\frac 1 2\sum_{i,j} V(\br_i-\br_j), \quad V(\br)=A\exp \brt{-\frac 1 {1- \pnt{r/\sigma }^2} } ,
\end{equation}
for which Fig.~\ref{fig:MIPS} shows that AOUPs exhibit 
Motility-Induced Phase Separation (MIPS)~\cite{Tailleur:08,Cates:15},
extending this phenomenon beyond the reported cases of
RTPs~\cite{Tailleur:08,Solon:15c} and
ABPs~\cite{Fily:12,Redner:13,Brader:15}. Our analytical results,
however, are valid beyond this example, and hold for general
potentials and dimensions. Introducing the velocities $\bp_i$, the
dynamics~\eqref{eq:AOU} become
\begin{equation}\label{eq:dyn}
  \tau \dot { \bp}_i = -  \bp_i - \pnt{ 1 + \tau  \bp_k \cdot \nabla_k } \nabla_i \Phi - \sqrt{2T} \boldsymbol{\eta}_i ,
\end{equation}
where the mobility $\mu$ is set to one. Here and in what
follows, repeated indices are implicitly summed over. 

We have introduced $D\equiv \mu T$ in Eq.~\eqref{eq:dyn} to make the
equilibrium limit $\tau=0$ transparent. Surprisingly, it suffices to
take either $\tau=0$ in the r.h.s. or in the l.h.s. of
Eq.~\eqref{eq:dyn} to map AOUPs onto (different) equilibrium
dynamics. Suppressing the non-linear damping in the r.h.s. indeed maps
Eq.~\eqref{eq:dyn} onto an underdamped Kramers-Langevin
equation. Conversely, neglecting $\tau \dot \bp_i$ in the
l.h.s. corresponds to the Unified Colored Noise Approximation~\cite{Hanggi:87,Maggi:15b} which has been shown to satisfy
detailed balance~\cite{Maggi:15b}. Here, we propose to determine
perturbatively the steady-state of AOUPs in the small $\tau$ limit,
retaining both contributions of $\tau$ in
Eq.~\eqref{eq:dyn}. Rescaling time as $t= \sqrt\tau \tilde t$ and
introducing rescaled velocities $ \tilde\bp_i= \sqrt\tau \bp_i$, the
steady-state distribution satisfies ${\cal L} P ( \{\br_i, \tilde \bp_i \})=0$, with
\begin{equation}
  \begin{aligned}
    {\cal L}&=- \tilde p_{i \alpha}  \frac{\partial}{\partial r_{i \alpha}} + \frac{1}{\sqrt\tau} \frac{\partial}{\partial \tilde  p_{i \alpha}} \left[ \tilde  p_{i \alpha}+ \tau \frac{\partial^2 \Phi}{\partial r_{i \alpha}r_{j \beta}}\tilde  p_{j \beta} \right]\\
  & \qquad +\frac{\partial }{\partial \tilde p_{i \alpha}} \frac{\partial \Phi}{\partial r_{i \alpha}}+\frac T {\sqrt\tau}\frac{\partial^2}{\partial \tilde p_{i \alpha}^2} .
\end{aligned}\label{eq:EO}
\end{equation}
Using the ansatz
\begin{equation}
  P\propto \exp\left[-\frac{\Phi}{T}-\frac{\tilde \bp_i^2}{2 T} + \sum_{n=2}^\infty \tau^{n/2} \psi_n(\{\br_i,\tilde \bp_i\}) \right] ,
\end{equation}
we obtain a set of equations at every order in $\sqrt\tau$ which can
be solved recursively to yield
\begin{equation}
  \begin{aligned}
    P&\propto {\rm e}^{-\frac{\Phi+ \tilde \bp_i^2/2}{T}} \Big \{ 1 - \f{\tau}{2T} \brt{ \pnt{\nabla_i \Phi}^2 +  (\tilde\bp_i \cdot \nabla_i)^2 \Phi -3 T \nabla_i^2 \Phi }
\\
& + \frac{\tau^{3/2}}{6T} (\tilde\bp_i \cdot \nabla_i)^3 \Phi -\frac{\tau^{3/2}}2 (\tilde\bp_i \cdot \nabla_i) \nabla_j^2 \Phi  + \cO\pnt{\tau^2} \Big \} .
\end{aligned}\label{eq:Ps}
\end{equation}
The distribution of positions can then be deduced by
integrating~\eqref{eq:Ps} over velocities; we define an effective
potential $\tilde \Phi$ by analogy with the Boltzmann measure:
$P(\{\br_i\})\propto \exp(- \beta\tilde\Phi)$, with $\beta\equiv
T^{-1}$ and
\begin{equation}\label{eq:tPhi}
\tilde \Phi \equiv \Phi + \tau [ (\nabla_i \Phi )^2 / 2 - T \nabla_i^2 \Phi ] + \cO\pnt{\tau^2} .
\end{equation}

In the limit of vanishing $\tau$, one recovers the standard
Maxwell-Boltzmann distribution. The joint distribution of position and
velocities~\eqref{eq:Ps} beyond this regime is our first important
result. First, it shows how, for finite $\tau$, positions and
velocities are correlated, in agreement with the UCNA
approximation~\cite{marconi:2016} but at contrast to thermal
equilibrium where the energy can be separated between kinetic and
potential parts. In particular, this leads to a modified equipartition
theorem:
\begin{equation}
\avg{ \tilde \bp_{i\alpha}^2 } = T  - \tau \big \langle  \pnt{\nabla_i \Phi}^2 \big\rangle_\B +\cO(\tau^2), 
\end{equation}
where the average $\langle \cdots \rangle_\B$ is taken with respect to
the Boltzmann weigth $P_\B$. Second, the effective potential $\tilde
\Phi$ predicts that repulsive pairwise potentials lead to effective
attractive interactions, consistently with other approximation
schemes~\cite{Brader:15,Maggi:15b}. This explains why purely repulsive
interactions can trigger MIPS. Note also how a pairwise potential
leads to effective three-body interactions through the term
$(\nabla_i\Phi)^2$. At this stage, our controlled expansion allows us
to describe the static properties of AOUPs in terms of an effective
Boltzmann weight~\eqref{eq:tPhi}. Interestingly, for the evolution
operator~\eqref{eq:EO}, the asymmetry in $\tilde {\bp_i}$ of the
steady-state measure~\eqref{eq:Ps} implies that the dynamics is
out-of-equilibrium~\cite{Gardiner:1985}. This asymmetry is not
captured by UCNA approximation~\cite{marconi:2016} which cannot
describe the non-equilibrium properties of AOUPs.

To better measure the degree of irreversibility of the dynamics, we
derive its entropy production rate $\sigma$~\cite{Lebowitz}. It is
obtained by comparing the probability weights associated with a given
trajectory and its time-reversed counterpart, respectively denoted by
$\cP$ and $\cP^\R$:
\begin{equation}
  \sigma \equiv \underset{t\to\infty}{\lim}  \frac 1 t \ln \frac\cP {\cP^\R} .
\end{equation}
To keep the scaling in $\tau$ explicit, we work for now with the
rescaled variables $\tilde t$ and $\tilde \bp_i$ and use the fact
that $\sigma$ is intensive in time. Using standard path-integral
formalism~\cite{Onsager:53}, the trajectory weight can be written as
$\cP[\{\br_i(\tilde t)\}]\sim \exp(- {\cal S}[\{\br_i(\tilde t)\}])$
with
\begin{equation}\label{eq:action}
{\cal S}=\frac{\sqrt\tau}{4T}\int_0^{\tilde t} {\rm d}u \Big[\dot {\tilde \bp}_i +\frac{\tilde \bp_i}{\sqrt\tau}+(1+\sqrt\tau\tilde \bp_k\cdot \nabla_k)\nabla_i \Phi\Big]^2 .
\end{equation}
The time-reversed trajectories are then given by $t^\R=-t$,
$\br_i^\R(t) \equiv \br_i(-t)$ and $\bp_i^\R(t) \equiv - \bp_i(-t)$ so
that ${\cal P}^\R$ is simply obtained by injecting these expressions
in~\eqref{eq:action}.  The entropy production rate is thus given by
$\sigma\sim \delta {\cal S}/t$ where $\delta {\cal S}$ is the
difference between the forward and backward actions~\footnote{Note
  that we did not carefully specify the time-discretization of the
  Langevin equation~\eqref{eq:dyn} since it does not affect
  $\sigma$.}. All in all, the entropy production rate
reads
\begin{equation}\label{eq:sig}
  \sigma=-\underset{\tilde t\to\infty}{\lim}\frac{ \sqrt \tau}{T\tilde t} \int_0^{\tilde t} {\rm d}u ({\dot {\tilde \bp}_i}\cdot \nabla_i)({\tilde \bp}_j \cdot \nabla_j) \Phi=\frac{\sqrt\tau}{2T} \langle (\tilde \bp_i \cdot \nabla_i)^3 \Phi \rangle ,
\end{equation}
where the last equality follows from integrating by parts and using
the equality between time and ensemble averages in
steady-state~\footnote{Note that temporal boundary terms, appearing
  for instance after integration by parts, do not contribute to
  $\sigma$ thanks to the $1/t$ prefactor}. Interestingly, the entropy
production rate exactly vanishes when $\Phi$ is quadratic in the
particle displacements, hence showing that AOUPs are in this case an
equilibrium model. Their steady-state is however not the Boltzmann
measure $P_{\B}$, which explains the difficulty of defining a
temperature in this case~\cite{Szamel:14}. As a result, the
anharmonicity of the potential acts as a control parameter for the
nonequilibrium nature of AOUPs.

The entropy production rate can also be computed in the small
$\tau$ limit, using the stationary distribution~\eqref{eq:Ps} to
evaluate the correlation function appearing in
Eq.~\eqref{eq:sig}. Going back to the initial variables, the entropy
production rate is given by
\begin{equation}\label{eq:sig2}
   \sigma = \f{ T\tau^{2}}{2} \Big\langle ({\nabla}_i {\nabla}_j {\nabla}_k \Phi)^2\Big\rangle_\B  + \cO\pnt{\tau^3} .
\end{equation}
The first non-vanishing contribution to $\sigma$ comes from the
$\tau^{3/2}$ correction in the steady-state measure~\eqref{eq:Ps}. At
order $\tau$, we thus have a non-Boltzmann steady-state given by the
first line of~\eqref{eq:Ps}, or equivalently by~\eqref{eq:tPhi} in
position space, with a vanishing entropy production rate. In such a
regime, the AOUPs are effectively a non-thermal \textit{equilibrium}
model which is the central result of this letter.

Let us now discuss the practical consequences of this effective
equilibrium dynamics. Oscillatory shear experiments have become an
increasingly standard procedure to sample the microrheology of active
systems~~\cite{Mizuno,Wilhelm, Robert, Ahmed:15}. In this context,
the violation of the equilibrium FDT has proven a natural measure of
the distance to equilibrium~\cite{Loi2008PRE,Loi2011SM,Levis:15}. Let
us consider that an external operator perturbs the dynamics by
applying a small constant force ${\bf f}_j$ on the particle $j$, hence
modifying the potential $\Phi$ as $\Phi \to \Phi - {\bf f}_i \cdot
\br_i $. We define the response function $R$ as
\begin{equation}
R_{i\alpha j\beta}(t,s) \equiv \left. \f{\delta \avg{ {r}_{i\alpha}(t)} }{\delta {f}_{j\beta}(s)} \right|_{ {\bf f} = { 0} } .
\end{equation}
Following standard procedures~\cite{Cugliandolo:11}, we can use the
dynamic action formalism and the fact that $\delta {\cal P}=-\delta {\cal S}.
{\cal P}$ to rewrite the response as
\begin{equation}
  R_{i\alpha j\beta}(t,s) = -\left. \left \langle r_{i\alpha}(t) \frac{\delta {\cal S}}{\delta {f}_{j\beta}(s)} \right|_{ {\bf f} = {0} } \right\rangle .
\end{equation}
The perturbed dynamics of the AOUPs is readily given by
\begin{equation}\label{eq:dynp}
  \tau \dot {\bp}_i = - \bp_i - \pnt{ 1 + \tau \bp_k \cdot \nabla_k } \nabla_i \Phi + {\bf f}_i+\tau \dot{\bf f}_i - \sqrt{2T} \boldsymbol{\eta}_i ,
\end{equation}
so that the dynamical action ${\cal S}$ becomes
\begin{equation}
  {\cal S}=\frac{1}{4T} \int_0^t {\rm d}u \left[ \Big(1+\tau\frac{\dd}{\dd u}\Big) (\bp_i+\nabla_i \Phi-{\bf f}_i)  \right]^2 .
\end{equation}
The response function is then given by
\begin{eqnarray}\label{eq:resp}
  R_{i\alpha j\beta}(t,s)&=&\left(1-\tau^2 \frac{\dd^2}{\dd t^2}\right) \left[-\frac 1 T \frac{\dd}{\dd t}  \langle r_{i\alpha}(t) r_{j\beta}(s)\rangle\right.\\ 
  & & \left. + \frac 1 {2T} (\langle r_{i\alpha}(t) \nabla_{j\beta}\Phi|_{t=s}\rangle - \langle r_{i\alpha}(s) \nabla_{j\beta}\Phi|_t\rangle )\right]\nonumber .
\end{eqnarray}
In the effective equilibrium regime, the vanishing entropy production
tells us that the dynamics is symmetric under time reversal so that
the second line of Eq.~\eqref{eq:resp} vanishes and the response
function finally reads:
\begin{equation}\label{eq:respf}
  R_{i\alpha j\beta}(t,s)= -\frac 1 T \frac{\dd}{\dd t}  \big\langle r_{i\alpha}(t) r_{j\beta}(s) +  \tau^2 p_{i\alpha}(t) p_{j\beta}(s)\big\rangle .
\end{equation}
We have thus derived a generalized FDT which holds in the small $\tau$
limit where the AOUPs are effectively in equilibrium, though not with
respect to the Boltzmann measure $P_{\B}$. This explains the atypical
form of the correlation function entering, which involves the position
autocorrelation function, as in thermal equilibrium, along with the
velocity autocorrelation function. Note that, as in equilibrium, this
FDT is completely independent of the interaction potential $\Phi$, so
that it should be measurable without knowledge of the intimate details
of particle interactions.

\begin{figure}
  \includegraphics[width=.6\columnwidth]{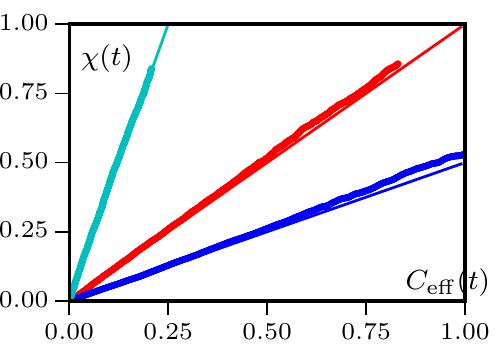}
  \caption{Parametric plot between the susceptibility $\chi(t)$ and
    the correlation function $C_{\rm eff}(t)=\langle
    x_i(t)(x_i(t)-x_i(0))+\tau^2 \dot x_i(t)(\dot x_i(t)-\dot x_i(0))
    \rangle$ for $N$ AOUPs interacting via the
    potential~\eqref{eq:repul}. The particles experience a stiff
    harmonic potential when they try to exit a box of linear size
    $L$. Parameters: $L=30$, $N=720$, $\tau=0.01$, $A=20$. Blue, red
    and cyan dots correspond to $T=2,1,0.25$ and the solid line
    correspond to the theoretical prediction~\eqref{eq:fdr}.}
  \label{fig:FDT}
\end{figure}

To test whether a finite $\tau$ regime exists where our generalized
FDT can indeed be measured, we consider a perturbation $\Phi \to \Phi
-f \varepsilon_i x_i$ where $\varepsilon_i$ is a random variable equal to
$\pm 1$ with equal probability~\cite{Levis:15}. We measure the susceptibility $\chi(t)
\equiv \int_0^t \dd s R_{ixix}(t,s) / N$ in simulations of AOUPs
interacting with the repulsive potential~\eqref{eq:repul}. Our
modified FDT predicts that
\begin{equation}\label{eq:fdr}
  N T \chi(t) = \avg{ [x_i(0)-x_i(t)]x_i(t)} + \tau^2 \avg{ [\dot x_i(0)-\dot x_i(t)]\dot x_i(t)} ,
\end{equation}
which is shown to be valid at small $\tau$ in
Figure~\ref{fig:FDT}a. 

Note that an entropy production rate $\sigma$ of order $\tau^2$ means
that trajectories of length $\propto \tau^{-2}$ lead to an overall
entropy production of order one. Since we are working in the
small-$\tau$-but-finite-$D$ limit, diffusive equilibration times
$\ell^2/D$ remain of order one, which legitimates the claim of an
effective equilibrium regime. Nevertheless, we expect our FDT to break
down in the long time limit. 

To get more physical insight into our effective equilibrium regime and
its breakdown as $\tau$ increases, let us now discuss the energetics
of AOUPs. Active matter is traditionnaly regarded as a non-equilibrium
medium because injection and dissipation of energy are
uncorrelated. Indeed the former stems from the conversion of some form
of stored energy while the latter results from the friction with the
surrounding medium. Consequently, fluctuations and dissipations are
not constrained by any form of Stokes-Einstein relations. For driven
Langevin processes, the non-equilibrium nature of the dynamics can be
measured as a mean heat transfer between particles and
thermostat~\cite{Sasa,Sekimoto}. This leads to a standard definition
of dissipation $J$ as the imbalance between the power injected by the
thermal noise and the one dissipated \textit{via} the drag force. This
definition furthermore provides an energetic interpretation of the
entropy production since $J=T \sigma$. A naive
generalization of this reasoning to AOUPs would lead to the definition
of dissipation through
\begin{equation}
  J = \mu^{-1} \avg{ \bp_{i} \cdot ( \bp_{i} - \bv_i ) } .
\end{equation}
It is however straightforward to see that $J=\avg{\bp_i \cdot \nabla_i
  \Phi}=\dd\avg{ \Phi} / \dd t$ which necessarily vanishes in
steady-state.

The breakdown of detailed balance for AOUPs is indeed not linked to a
mean heat flux extracted from an equilibrated bath but from the
apparent lack of generalized FDT between damping and fluctuations
in~\eqref{eq:AOU}. To get more insight on the entropy production rate
$\sigma$, we remark that this dynamics is equivalent to
\begin{equation}\label{eq:correl}
  K * \dot \br_i = {\boldsymbol \xi}_i -\mu K * \nabla_i \Phi ,
\end{equation}
where $K(t)=[1-\tau^2 (\dd/{ \dd t})^2 ]\delta(t) $, $*$ denotes time
convolution, and we have introduced the noise term ${\boldsymbol
  \xi}_i \equiv K * \bv_i$. The lhs of~\eqref{eq:correl} corresponds
to the damping of a visco-elastic fluid with memory kernel $K$. The
first term on the rhs is a fluctuating force whose variance is:
\begin{equation}
  \avg{ \xi_{i\alpha}(t) \xi_{j\beta}(0)} = \delta_{ij}\delta_{\alpha\beta} K(t) ,
\end{equation}
since by definition $(K*\Gamma)(t)=\delta(t)$. The
damping and fluctuating forces appearing in~\eqref{eq:correl} thus
satisfy a generalized Stokes-Einstein relation~\cite{Kubo}. They
correspond to the connection of particles with an
\textit{equilibrated} visco-elastic bath, for which the standard
definition of the dissipation applies:
\begin{equation}\label{eq:diss}
  {\cal J} = \mu^{-1} \avg{ \bp_{i} \cdot ( K * \bp_{i} - {\boldsymbol \xi}_i ) } .
\end{equation}
From there, simple algebra shows that ${\cal J} = T \sigma$, which
yields a physical interpretation to $\sigma$ as the dissipation in an
equilibrated bath for the dynamics~\eqref{eq:correl}.

Interestingly, this shows that the breakdown of detailed balance in
AOUPs can be seen equivalently as resulting from a lack of generalized
Stokes-Einstein relation between damping and fluctuations or from the
fact that $K * \nabla_i \Phi$ is not a conservative force. In this
second interpretation, the entropy production rate now has a standard
energetic interpretation. The existence of an effective equilibrium
regime for small $\tau$ is then due to the fact that $K * \nabla_i
\Phi$ behaves as a conservative force $\nabla_i \tilde \Phi$ in this
limit. The dynamics~\eqref{eq:correl} with $K * \nabla_i \Phi$
replaced by $\nabla_i \tilde \Phi$ can be regarded as a dynamical
equilibrium approximation of AOUPs; one indeed checks, for instance,
that $\langle\tilde \Phi\rangle - \langle K*
\Phi\big\rangle=\cO(\tau^2)$ or that our generalized FDT corresponds
to perturbing this equilibrium dynamics as $\tilde \Phi \to \tilde
\Phi - \br_i \cdot ( K * {\bf f}_i )$.

In this article we have thus shown that, as their persistence time
increases, Active Ornstein-Uhlenbeck Particles do not immediately
leave the realm of equilibrium physics. At short persistent time, they
behave as an equilibrated visco-elastic medium with effective
Boltzmann weight $P\propto \exp(-\beta \tilde \Phi)$ which differs
from the thermal equilibrium $P_\B\propto \exp(-\beta \Phi)$. In
this regime, the fact that repulsive forces lead to effective
attractive interactions can directly be read in $\tilde \Phi$. Beyond
this static result, the existence of an effective equilibrium regime
enforces a generalized fluctuation dissipation theorem, akin to its
thermal counterpart though different correlators are involved. The
breakdown of this FDT for larger persistence times can be linked to a
non-zero entropy production rate whose expression we have computed
analytically. Last, we have shown how to extend the notion of
dissipation to understand the breakdown of detailed balance in AOUPs.

Most of the results presented in this letter have been derived for the
particular choice of noise correlator $\Gamma(t) = D \ee^{-|t|/\tau} / \tau$. Many of our results, such as the discussion on dissipation, however extends to more general
correlators. Furthermore, it has recently been shown that static
approximations derived for the steady-state of AOUPs capture very well
the physics of ABPs~\cite{Brader:15}. It would thus be very
interesting to know whether our effective equilibrium regime also
extends to this system. More generally, our study suggests that when
systems are driven out of thermal equilibrium by the conversion of
some form of stored energy, an effective equilibrium regime may remain
when the drive is moderate. This would be a first step towards a
thermodynamics of Active Matter.

\textit{Acknowledgements.} We thank M. Baiesi, JL Barrat, E. Bertin,
O. Grossein, C. Maggi, R. di Leonardo, G. Szamel, and for interesting
discussions. JT was supported by ANR project Bactterns. CN was
supported by EPSRC grant Nr. EP/J007404.

\bibliographystyle{apsrev4-1}
\bibliography{note-ref}

\end{document}